\documentstyle[prb,aps,twocolumn,psfig]{revtex}

\begin{document}
\wideabs{
\title{Spectral shapes of solid neon}
\author{Gaia Pedrolli\cite{e-GP}, Alessandro Cuccoli\cite{e-AC},
        Alessandro Macchi\cite{e-AM}, Valerio Tognetti\cite{e-VT}}
\address{Dipartimento di Fisica dell'Universit\`a di Firenze
         and Istituto Nazionale di Fisica della Materia (INFM),\\
         Largo E. Fermi~2, I-50125 Firenze, Italy}
\author{Ruggero Vaia\cite{e-RV}}
\address{Istituto di Elettronica Quantistica
         del Consiglio Nazionale delle Ricerche,
         via Panciatichi~56/30, I-50127 Firenze, Italy,\\
         and Istituto Nazionale di Fisica della Materia (INFM).}
\date{\today}
\maketitle
\begin{abstract}
We present a Path Integral Monte Carlo calculation of the first
three moments of the displacement-displacement correlation functions
of solid neon at different temperatures for longitudinal and
transverse phonon modes. The Lennard-Jones potential is considered.
The relevance of the quantum effects on the frequency position of
the peak and principally on the line-width of the spectral shape
is clearly pointed out. The spectrum is reconstructed via a
continued fraction expansion; the approximations introduced using
the effective potential quantum molecular dynamics are discussed.
\end{abstract}
} 

Rare gas solids (RGS) are the simplest real systems in which we can 
study lattice vibrations. Argon and the heavier RGS can be well 
approximated as a set of harmonic oscillators at lowest temperatures,
while the classical behaviour is reached before the melting point.
Quantum corrections on the thermodynamic quantities, like
kinetic energy and specific heat, can be taken into account by
means of the effective potential approach (EP) \cite{review_JPCM} 
only for small quantum coupling $g$, ($g<0.25$) defined as the ratio 
between the characteristic frequency and the strength of the binding 
potential. For neon ($g=0.694$) anharmonic effects are present even for
determining the ground state \cite{Ne_Kin_en,Acocella2}. Indeed,
precise calculations of kinetic energy, to be compared with accurate
experiments done by Deep Inelastic Neutron Scattering (DINS) \cite{Glyde},
needed a rather sophisticated Path Integral Monte Carlo (PIMC) 
computation, being the EP approach inadequate. These results
show that quantum effects are very important also at rather high
temperatures \cite{Ne_Kin_en}.

Information about phonon dynamics is given by spectral shape,
namely the space and time Fourier transform of the (symmetrized)
displacement-displacement correlation function:
\begin{equation}
{\cal S}_{\rm S}^{\alpha\beta}({\bf k},\omega) = \frac{1}{2\pi}
\sum_{\bf r} \int \! dt \; C^{\alpha\beta}({\bf r},t) \, 
\exp\left[i({\bf k} \cdot {\bf r} -\omega t)\right]\,\label{SCF}
\label{SOM}
\end{equation}
with
\begin{equation}
C^{\alpha\beta}({\bf r},t) = 
\left[\left\langle x^\alpha_{{\bf i}+{\bf r}}(t) x^\beta_{\bf i}(0)
\right\rangle + 
\left\langle x^\alpha_{{\bf i}+{\bf r}}(0) x^\beta_{\bf i}(t) 
\right\rangle \right].
\label{C(t)}
\end{equation}
$x_{\bf i}^\alpha$ is $\alpha$-th component of the displacement 
of the ${\bf i}$-th atom from its equilibrium position.
Even though this quantity has been investigated since long time 
\cite{Maradudin}, complete information is not available for
various rare gas solids and in particular for neon. 
The perturbative many-body approach can
give frequency and lifetime of phonons only at low temperatures.
Classical molecular dynamics (CMD) can describe the behaviour of argon 
and krypton at highest temperatures, but it is no more valid for 
lower temperatures or stronger coupling. As shown in the following,
the spectra of neon present significant quantum effects up to the 
melting point and the high quantum coupling prevents us to use the
EP method in the entire temperature range.

We approach the calculation of the spectra of neon as given in equation
(\ref{SOM}), by PIMC, evaluating the first three even frequency moments,
\begin{equation}
\left\langle \omega^{2n}\right\rangle_{\bf k}^{\alpha\beta} =
\int_{-\infty}^\infty \! \! d\omega \; \omega^{2n} 
{\cal S}_{\rm S}^{\alpha\beta}
({\bf k} ,\omega) ,
\label{MOM}
\end{equation}
while the odd moments vanish for symmetry reasons. 
As it is well known, this involve the PIMC calculation of static 
correlations obtained by multiple commutators of 
$x^\alpha_{\bf k}(t) = N^{-1/2} \sum_{\bf i} \exp(i {\bf k} \cdot
{\bf i} ) x^\alpha_{\bf i}(t)$ with the Hamiltonian:
\begin{equation}
\left\langle\omega^{2n}\right\rangle_{\bf k}^{\alpha\beta} =
\frac{1}{2} \left\langle 
\frac{d^n x^\alpha_{\bf k}}{dt^n}\bigg|_{0}
\frac{d^n x^\beta_{-{\bf k}}} {dt^n} \bigg|_{0} +
\frac{d^n x^\alpha_{-{\bf k}} }{dt^n} \bigg|_{0}
\frac{d^n x^\beta_{\bf k}}{dt^n} \bigg|_{0} \right\rangle,
\label{Mom}
\end{equation}
with the derivatives taken at $t=0$.
Here we will refer only to ${\bf k} = 2 \pi/a_0 (1,0,0)$
for which one longitudinal and two degenerate
transverse modes are present, so that we omit 
polarization indexes. 

When the spectra are sufficiently narrow, the normalized second
moment, $\delta_{1{\bf k}}$ 
and the irreducible part of fourth moment,
$\delta_{2{\bf k}}$:
\begin{equation}
\delta_{1{\bf k}} =
\frac{\left\langle\omega^2\right\rangle_{\bf k}}
{\left\langle\omega^0\right\rangle_{\bf k}}
\,\;\;;\;\;\delta_{2{\bf k}} =
\frac{\left\langle\omega^4\right\rangle_{\bf k}}
{\left\langle\omega^2\right\rangle_{\bf k}} -
\delta_{1{\bf k}},
\label{DELTA}
\end{equation}
can be directly related with the peak position and width of the 
spectra \cite{LoveseyMeserve}.

A reconstruction of the spectra can be done by the continued
fraction expansion of the Laplace transform of the normalized
correlation function,
\begin{equation}
\Xi_0({\bf k},z) =\int_0^\infty \!\! dt \; 
\frac{C({\bf k},t)}{C({\bf k},0)} \;
{\rm e}^{-zt}\,\label{Xi}
\end{equation}
 with
 \begin{equation}
{\cal S}_{\rm S}({\bf k},\omega) = \Re
\Xi_0({\bf k},z=i\omega) \; C({\bf k},t=0).\label{Xii}
\end{equation}
The continued fraction expansion can be stopped at the third stage 
with a suitable termination $\Xi_2({\bf k},z)$
\begin{equation}
\Xi_0({\bf k},z) \simeq \frac{1}{\pi} \,
\frac 1{z+\displaystyle\frac {\delta_{1{\bf k}}}
{z+\delta_{2{\bf k}}\Xi_2({\bf k},z)}}\,\,.\label{contfraz}
\end{equation}
In this paper, we shall present some of these spectra, showing the
validity of the approach. Neutron scattering data are not available
up to now and we suggest and discuss here the possibility to 
perform such an experiment.

We have considered samples of solid neon, with 256 atoms interacting
through a (12-6) Lennard-Jones pairwise potential,
\begin{equation}
V(r)=4\epsilon\left[\left(\frac{\sigma}{r}\right)^{12}-
\left(\frac{\sigma}{r}\right)^6\right],
\label{LJ}
\end{equation}
and periodic boundary conditions.
The dynamic interaction is limited to the 12 nearest neighbours, 
while for the outer shells the static approximation is used.
For every temperature we resorted to three different
Trotter numbers: $P=8,16,24$. For each of them 16 simulation runs
of $100\,000$ steps per particle were performed,
plus $20\,000$ steps per particle for initial thermalization.
The density was adjusted in order to get a practically vanishing
pressure (the pressure is always less than $15$ atm).
The parameters of the Lennard-Jones potential are taken
as $\epsilon=36.68$~K and $\sigma=2.787$~\AA~\cite{Neon-parameter}.
The melting temperature of neon at zero pressure is $24.5$~K.

Detailed explanations of the total procedure and results for the
first three even moments \cite{thesis}, at different temperatures for
longitudinal and transverse modes, will be presented in an extended
paper. Here we want to point out that when the order
of the frequency moment increases, more complicated static
correlations are involved. Moreover accurate Trotter extrapolations 
are in order and finite-size effects are more and more important. 
We have used the procedure introduced by us \cite{Ne_Kin_en,trick95} 
by which we correct the raw PIMC data, subtracting the exact
contributions of the harmonic part for finite $P$ and $N$, 
and adding both the exact harmonic results for infinite $P$ and $N$,
In this way, an accuracy of $0.2\%$ is reached for the zeroth moment,
which rises to $1\%$ for the fourth moment. 
This corresponds to a maximum uncertainty of $4\%$ for
$\delta_{2{\bf k}}$. Classical simulations were also done for
comparison, using both the classical and the effective potential.

The necessity to account for quantum effects by an ``exact'' method
like PIMC has been ascertained, especially for higher-order
moments. The fourth moment exhibits strong quantum effects at any
realistic temperature which cannot be approached by the EP method. 
In tables \ref{mom10_table} and \ref{mom20_table}
we report the first three even moments together with the analogous
ones obtained by classical and EP Monte Carlo calculations. 
The fourth moment is related to the width of the spectra 
(phonon lifetime) through the quantity
$\delta_{2{\bf k}}$. For narrow spectra ($\delta_{2{\bf k}}
\ll \delta_{1{\bf k}}$),
the phonon frequencies are 
$\omega_{\bf k}\sim \sqrt{\delta_{1{\bf k}}}$
while the phonon lifetimes are 
$\eta_{\bf k}\sim \sqrt{\delta_{2{\bf k}}}/2$.
These quantities, shown in table \ref{delta_table}, 
can be probed by inelastic neutron scattering.

\begin{figure}
\centerline{\psfig{bbllx=30mm,bblly=33mm,bburx=205mm,bbury=240mm,%
figure=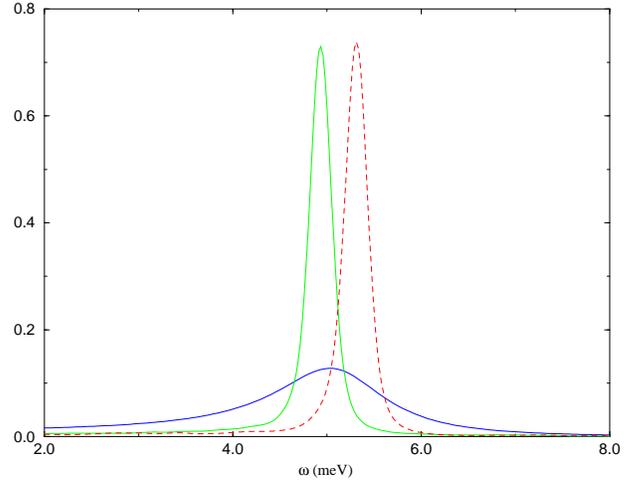,width=80mm,angle=270}}
\caption{Trasverse projection of the (normalized) space and time 
Fourier transform of the symmetrized displacement-displacement
correlation function,
${\cal S}_{\rm S}({\bf k},\omega)/C({\bf k},t=0)$.
at $T = 10$~K $\rho=1.494$~g cm$^{-3}$ 
for ${\bf k} = 2 \pi /a_0(1,0,0)$ and zero pressure.
The dashed line refers to CMD results,
the long-dashed one to EPMD 
results, and the solid line to results obtained via continued
fraction expansion and PIMC evaluation of frequency moments.
\label{10k_tr_fig} }
\end{figure}

\begin{figure}
\centerline{\psfig{bbllx=30mm,bblly=33mm,bburx=205mm,bbury=240mm,%
figure=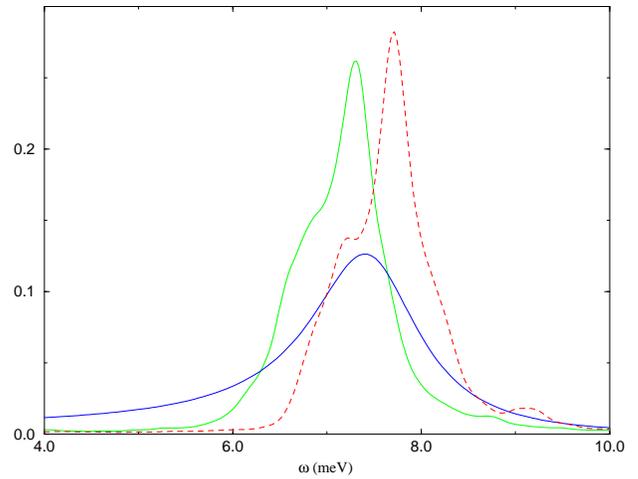,width=80mm,angle=270}}
\caption{The same as in figure \protect\ref{10k_tr_fig} for the longitudinal 
projection of the (normalized) space and time 
Fourier transform of the symmetrized displacement-displacement
correlation function,
${\cal S}_{\rm S}({\bf k},\omega)/C({\bf k},t=0)$,
at $T = 10$~K, $\rho=1.494$~g cm$^{-3}$ for 
${\bf k} = 2 \pi /a_0(1,0,0)$ 
\label{10k_long_fig} }
\end{figure}

Finally, we have calculated the spectra with the available
$\delta_{1{\bf k}},\delta_{2{\bf k}}$ by means of the continued
fraction (\ref{contfraz}) and with a suitable ``Gaussian
termination'' \cite{TomitaTomita}: 
$ \Xi_2({\bf k},t)=\exp\left(-\Gamma t^2\right)\,$.
The parameter $\Gamma$, is determined by the insight on the
corresponding spectra obtained by CMD\cite{spec_shape_LJ_ch}.

As an example, some spectra are shown in figures \ref{10k_tr_fig}
and \ref{10k_long_fig}. In these
figures we report also the similar spectra,  obtained by 
EP molecular dynamics (EPMD) \cite{Voth}.

We therefore can conclude that:
\begin{description}
\item{{i)}} Quantum effects in solid neon are relevant at all
temperatures so that neon cannot be approached by classical models.
\item{{ii)}} The evaluation of the spectral width requires 
a particular care and a fully quantum treatment of the fourth moment. 
EPMD, as observed in \cite{GiachettiTognetti97},
can correctly give the peak position only because it reproduces
just the short-time behaviour and consequently the second moment; 
using this method the phonon damping is therefore calculated 
by considering classical processes only. 
This is not sufficient for giving a good description of the spectra 
as on the other hand our approach is expected to do.
\end{description}
We conclude suggesting new accurate neutron scattering experiments
in order to investigate quantum effects in solid neon.
In particular our results on zero moment can be tested measuring
the integrated intensity, while the afore-mentioned features of
the line shapes can be directly compared with experimental spectra. 

\medskip
We would like to thank Prof. S. W. Lovesey for useful discussions. 
One of us (V. T.) wants also to thank him for hospitality in the very
stimulating atmosphere of the Rutherford Appleton Laboratory.

\newpage

\begin{table}
\caption{
Moments of ${\cal S}_{\rm S}({\bf k},\omega)$ for $T=10$~K at vanishing
pressure, for the wave vector ${\bf k} = 2 \pi/a_0 (1,0,0)$. They are
expressed in reduced units. $b = 2.787$~\AA~ and $\omega_0 = 0.289747$ meV. 
Classical second moments are exact since they are equal to the reduced
temperature. }
\begin{tabular}{@{}llll}
~ & \multicolumn{3}{c}{\bf Transverse} \\
~ & classical & EP & PIMC  \\
\tableline
$M_0/b^2\,(\times 10^{-4})$ & $8.427 \pm 0.005$ & $30.02 \pm 0.01$ &
$28.28 \pm 0.06$ \\
$M_2/(b^2 \omega_0^2)$ & 0.2726281 & $ 0.7027 \pm 0.0002$ 
& $0.788 \pm 0.002$ \\
$M_4/(b^2 \omega_0^4)$ & $98.13 \pm 0.05$ & $168.5 \pm 0.1$ 
& $314. \pm 3.$  \\
\tableline
\tableline
~& \multicolumn{3}{c}{\bf Longitudinal} \\
~ & classical & EP & PIMC  \\
\tableline
$M_0/b^2\,(\times 10^{-4})$ & $3.955  \pm 0.002$ &
$20.732  \pm 0.008 $ & $19.02  \pm 0.05 $ \\
$M_2/(b^2 \omega_0^2)$ &
0.2726281 & $0.9902 \pm 0.0003$ & $1.141 \pm 0.004$ \\
$M_4/(b^2 \omega_0^4)$ & 
$205.0 \pm 0.2$ & $467.3 \pm 0.3$ & $ 849. \pm 6.$ \\ 
\end{tabular}
\label{mom10_table}
\end{table}

\bigskip

\begin{table}
\caption{
Moments of ${\cal S}_{\rm S}({\bf k},\omega)$ for $T=20$~K at vanishing
pressure, for the wave vector ${\bf k} = 2 \pi/a_0 (1,0,0)$. They are
expressed in reduced units. $b = 2.787$~\AA~ and $\omega_0 = 0.289747$ meV.
Classical second moments are exact since they are equal to the reduced
temperature. }
\begin{tabular}{@{}llll}
~ & \multicolumn{3}{c}{\bf Transverse} \\
~ & classical & EP & PIMC \\
\tableline
$M_0/b^2\,(\times 10^{-4})$ & $20.92 \pm 0.02$ & $36.50 \pm 0.05$
 & $34.7 \pm 0.3$ \\
$M_2/(b^2 \omega_0^2)$ & 0.545256 & $ 0.8387 \pm 0.0006$ 
 & $0.870 \pm 0.002$ \\
$M_4/(b^2 \omega_0^4)$ & $182.3 \pm 0.3$ & $218.6 \pm 0.4$ 
 & $347. \pm 3.$ \\
\tableline
\tableline
~ & \multicolumn{3}{c}{\bf Longitudinal} \\
~ & classical & EP & PIMC  \\
\tableline
$M_0/b^2\,(\times 10^{-4})$ &
$9.67  \pm 0.01$ & $22.47  \pm 0.02 $ & $21.3  \pm 0.1 $ \\
$M_2/(b^2 \omega_0^2)$ & 
0.545256 & $1.1022 \pm 0.0006$ & $1.157 \pm 0.006$ \\
$M_4/(b^2 \omega_0^4)$ & 
$380.3 \pm 0.5$ & $560.1 \pm 0.3$ & $ 872. \pm 8.$ \\
\end{tabular}
\label{mom20_table}
\end{table}

\bigskip

\begin{table}
\caption{
$\protect\sqrt{\delta_1}$ and $\protect\sqrt{\delta_2}/2$ roughly
represent the peak position and the phonon life time, respectively. 
They are evaluated via PIMC simulations and they are expressed in meV. }
\begin{tabular}{@{}lllll}
~ & \multicolumn{2}{c}{\bf Transverse} &
\multicolumn{2}{c}{\bf Longitudinal} \\
T & $\sqrt{\delta_1}$ & $\sqrt{\delta_2}/2$ &
$\sqrt{\delta_1}$ & $\sqrt{\delta_2}/2$  \\ 
\tableline
10 & $4.83 \pm 0.02$ & $1.58 \pm 0.1$ &
$7.01 \pm 0.02$ & $1.78 \pm 0.06$ \\
20 &
$4.57 \pm 0.02$ & $1.74 \pm 0.1$
& $6.76 \pm 0.02$ & $2.07 \pm 0.06$ \\
\end{tabular}
\label{delta_table}
\end{table}

\end{document}